# Second sound in nuclear interaction

Magdalena Pelc
Institute of Physics Maria Curie –Sklodowska Universiy
Lublin Poland

Miroslaw Kozlowski*
Institute of Physics Warsaw University, Warsaw, Poland

* Corresponding author: e-mail: miroslawkozlowski@aster.pl

Abstract: Following the method developed in monograph : From quarks to bulk matter Hadronic Press 2001the master equation for second sound in nuclear matter is obtained and solved. The velocity of second sound is calculated

1. Introduction

The present paper is devoted to study of second sound propagation in nuclear matter The velocity of heat propagation is calculated and the value $v_S = v_F/\sqrt{3}$ is obtained. Assuming the relaxation-time approximation, the diffusivity $D$ and viscosity $\eta$ are obtained. The second sound master equation is used to study the heating of two interacting nuclear slabs. The obtained solution of the equation describes two modes of heat transfer in nuclear matter: (i) ballistic propagation of the temperature pulse and (ii) heat diffusion. The temperature of the two interacting nuclear slabs is calculated

2. Theory

Let us consider the propagation of the heat in a nuclear medium. We will describe the nucleus as a Fermi gas of nucleons When the nucleus is subjected to a heat pulse, e.g. during nuclear reaction the propagation of the pulse inside the nucleus can be described by a nonlocal Fourier law [1]

$$\vec{q} = -\int_{-\infty}^{t} K(t-t')\nabla T(x',t')dt' \tag{1}$$

where $q(t)$ denotes energy flux density, $T$ is the temperature (in MeV) and $K(t\text{-}t')$ denotes the memory function for thermal processes. The energy flux density $q(t)$ fulfils the nonlocal heat transport equation:

$$\frac{\partial}{\partial t}T(t) = \frac{1}{\rho c_V}\nabla^2 \int_{-\infty}^{t} K(t-t')T(t')dt', \tag{2}$$

where $\rho$ is the density of Fermi gas of nucleons and $C_V$ is the specific heat at constant volume.

For a system with very short memory

$$K(t-t') = K_1\delta(t-t'). \tag{3}$$

we obtain from eq.(2)

$$\frac{\partial}{\partial t}T(t) = \frac{1}{\rho c_V}K_1\nabla^2 T. \tag{4}$$

By comparing eq.(4) with the kinetic theory formulation of heat transport [8] we obtain for $K_1$

$$K_1 = D c_V \rho, \tag{5}$$

where $D$ is the diffusivity of the nuclear medium and $C_V$ is the specific heat at constant volume, $\rho$ denotes the density of nuclear matter.

Next, we consider a system with a very long memory function, i.e.

$$K(t-t') = K_2. \tag{6}$$

After substituting formula (6) into formula (2) we obtain

$$\frac{\partial}{\partial t} T = \frac{K_2}{\rho c_V} \nabla^2 \int_{-\infty}^{t} T(t') dt, \tag{7}$$

and finally

$$\frac{\partial^2 T}{\partial t^2} = \frac{K_2}{\rho c_V} \nabla^2 T. \tag{8}$$

In the following, we consider the intermediate form of the memory function:

$$K(t-t') = \frac{K_3}{\tau} \exp\left[-\frac{(t-t')}{\tau}\right], \tag{9}$$

where $\tau$ is the relaxation time for thermal phenomena. After substituting formula (9) into eq. (2) we obtain

$$c_V \frac{\partial^2 T}{\partial t^2} + \frac{c_V}{\tau} \frac{\partial T}{\partial t} = \frac{K_3}{\rho \tau} \nabla^2 T, \tag{10}$$

where

$$K_3 = D c_V \rho \tag{11}$$

as in formula (5). Considering formula (11) we obtain from formula (10)

$$\frac{\partial^2 T}{\partial t^2} + \frac{1}{\tau} \frac{\partial T}{\partial t} = \frac{D}{\tau} \nabla^2 T. \tag{12}$$

For a gas of nucleons we have for the diffusivity [2]

$$D = \frac{1}{3} v_F^3 \tau, \tag{13}$$

where $v_F$ is the Fermi velocity. Substituting formula (13) into formula (12) we obtain the following general equation for the second sound in Fermi gas

$$\frac{\partial^2 T}{\partial t^2} + \frac{1}{\tau} \frac{\partial T}{\partial t} = \frac{1}{3} v_F^3 \nabla^2 T. \tag{14}$$

Let us introduce the propagation velocity, $s$, for the heat pulse

$$s = \sqrt{\frac{1}{3}} v_F, \qquad (15)$$

then eq. (14) has the form:

$$\frac{1}{s^2} \frac{\partial^2 T}{\partial t^2} + \frac{1}{\tau s^2} \frac{\partial T}{\partial t} = \nabla^2 T. \qquad (16)$$

It is interesting to observe that for Fermi liquids one can define the sound velocity as [3]

$$v_S = \left( \frac{p_F^2}{3mm^*} (1 + F_0^S) \right)^{1/2}, \qquad (17)$$

where $m$ is the mass of the free fermion (e.g. the nucleon), $m^*$ denotes the effective mass of interacting fermions and $F_0^S$ is the dimensionless measure of the interaction strength in the Fermi system. In the limit of weak interactions, $m^* \to m$, $F_0^S \to 0$ The velocity of sound then tends towards $\sqrt{\frac{1}{3}} v_F$. In that case we observe for $v_s$,

$$v_S \to s \qquad (18)$$

and eq. (16) can be written in the form

$$\frac{1}{v_S^2} \frac{\partial^2 T}{\partial t^2} + \frac{1}{\tau v_S^2} \frac{\partial T}{\partial t} = \nabla^2 T. \qquad (19)$$

Eq. (19) is the damped wave equation for the propagation of the second sound pulse with propagation velocity $v_s$, which is the velocity of second sound in a nucleon Fermi gas. Assuming for the Fermi energy in the nucleus $E_F = 40$ MeV we obtain for the second sound velocity $v_s = 0.17\ c$, where $c$ is the velocity of light. Considering that for a Fermi gas one can define the diffusivity $D$ as:

$$D = \tau v_S^2, \qquad (20)$$

where for nuclear matter the relaxation time $\tau \sim 10^{-23}$ s = 3 fm/$c$ eq. (19) can be written as

$$\frac{1}{v_S^2} \frac{\partial^2 T}{\partial t^2} + \frac{1}{D} \frac{\partial T}{\partial t} = \nabla^2 T \qquad (21)$$

with $D = 8.7 * 10^{-2}$ fm·$c$. For a Fermi gas the viscosity $\eta$ is connected to diffusivity through formula

$$\eta = D\rho. \qquad (22)$$

Assuming $D = 8.7 * 10^{-2}$ fm·$c$ and $\rho = 135.57$ MeV/($c^2$fm$^3$) we obtain for the viscosity

$$\eta = 11{,}79\ \text{MeV/(fm}^2 c). \qquad (23)$$

The obtained value of bulk viscosity $\eta$ is in good agreement with presently adopted value of $\eta$ for nuclear matter, $\eta = 3 - 15$ MeV $c$ / fm$^2$

## 3. The second sound propagation in nuclear matter

Consider a cylindrical sample of nuclear matter with unit area which is heated at one end. The temperature at the other end of the sample is detected as a function of time. The purpose of this section is to discuss the solution of eq.(19) which will be relate the signal at the temperature pulse detector (TP) $T(l, t)$ to the input pulse $T(0, t)$. The solution of eq.(19) for a cylinder of infinite length is given by

$$T(x,t) = \frac{1}{2v_S} \int dx' T(x',0) \left[ e^{-t/2\tau} \frac{1}{t_0} \delta(t-t_0) + e^{-t/2\tau} \frac{1}{2\tau} \left\{ I_0\left(\frac{(t^2-t_0^2)^{1/2}}{2\tau}\right) + \frac{t}{(t^2-t_0^2)^{1/2}} I_1\left(\frac{(t^2-t_0^2)^{1/2}}{2\tau}\right) \right\} \right] \Theta(t-t_0) \quad (24)$$

where $t_0 = (x-x')/v_S$ and $I_0$ and $I_1$ are modified Bessel functions. We are concerned with the solution to eq.(21) when a nearly delta-function temperature pulse heats one end of the sample. Then at $t \sim 0$ the temperature distribution in the sample is

$$T(x,t) = \begin{cases} \Delta T_0 & \text{for } 0 < x < v_S \Delta t, \\ 0 & \text{for } x > v_S \Delta t. \end{cases}$$

With this $t = 0$ temperature profile, eq.(19) yields

$$T(l,t) = \tfrac{1}{2} \Delta T_0 e^{-t/2\tau} \Theta(t-t_0)\Theta(t_0+\Delta t - t)$$
$$+ \tfrac{1}{4} \Delta t \Delta T_0 e^{-t/2\tau} \left\{ I_0(z) + \frac{t}{2\tau} \frac{1}{z} I_1(z) \right\} \Theta(t-t_0), \quad (25)$$

where $z = (t^2-t_0^2)^{1/2}/2\tau$ and $t_0 = l/v_S$. The first term in this solution corresponds to ballistic propagation of the second sound damped by $\exp(-t/2\tau)$ across the sample. The second term corresponds to the propagation of the energy scattered out of the ballistic pulse by diffusion. In the limit $\tau \to \infty$ the ballistic pulse alone arrives at the detector. In the limit $\tau \to 0$ the ballistic pulse is completely damped and the second term takes an asymptotic form which is the solution to the conventional diffusion equation. In particular in this limit we have $z \to \infty$

$$I_0(z) \sim \frac{e^z}{(2\pi z)^{1/2}}, \qquad I_1(z) \sim \frac{e^z}{(2\pi z)^{1/2}}, \qquad (26)$$

so that the second term in eq.(25) becomes

$$T(l,t) \sim 2 \frac{\Delta t}{4\tau} \Delta T_0 \frac{e^{-t/2\tau} \exp\left[-(t^2 - t_0^2)^{1/2}/2\tau\right]}{\left[(\pi/\tau)(t^2 - t_0^2)^{1/2}\right]^{1/2}}. \qquad (27)$$

Now, for $t \gg t_0$, we can write $(t^2 - t_0^2)^{1/2} \sim t - \tfrac{1}{2}(t_0^2/t)$ and thus obtain

$$\lim_{\substack{t \gg t_0 \\ \tau \to 0}} T(l,t) = \Delta T_0 \frac{\Delta t}{(4\pi\tau t)^{1/2}} \exp\left(-\frac{t_0^2}{4t\tau}\right). \qquad (28)$$

The solution to eq.(21) when there are reflecting boundaries on the sample is the superposition of the temperature at $l$ from the heated end and from image heat sources at $\pm 2nl$. This solution is

$$T(l,t) = \sum_{i=1}^{\infty} \left[ \begin{array}{l} \Delta T_0 e^{-t/2\tau} \Theta(t-t_i)\Theta(t_i + \Delta t - t) \\ + \Delta T_0 \dfrac{\Delta t}{2\tau} e^{-t/2\tau} \left\{ I_0(z_i) + \dfrac{t}{2\tau} \dfrac{1}{z_i} I_1(z_i) \right\} \Theta(t - t_i) \end{array} \right], \qquad (29)$$

$t_i = t_0, 3t_0, 5t_0, \ldots = l/v_S$. For short times only the $i = 0$ term in the sum is important. Each source begins to contribute at $x = l$ at the time (1) proportional to its distance (in the ballistic limit) or (2) proportional to the square of its distance (in the diffusion limit). After sufficiently long times one obtains always the diffusion limit. At the time $t_1 \gg t_0$ all sources at a distance greater than $l_1$ given by

$$\left(\frac{l_1}{v_S \tau}\right)^2 \tau = \frac{(t_1)^2}{\tau} \qquad (30)$$

do not contribute to the temperature at $l$ at $t_1$. Each source which contributes to the temperature at $l$ contributes the same amplitude

$$A_i = \Delta T_0 \frac{\Delta t}{(\pi \tau t_1)^{1/2}}, \qquad (31)$$

so that the total heat at $l$ is

$$T(l,t) = \sum_{i=1}^{N} \frac{\Delta T_0 \Delta t}{(\pi \tau t_1)^{1/2}}, \qquad (32)$$

where the number of contributing sources $N$ is given by $(2N+1)l = l_1$ or $N \sim (v_S/l)[\frac{1}{2}(\tau t_1)]^{1/2}$ so that eq.(32) becomes

$$T(l,t) \sim N \frac{\Delta T_0 \Delta t}{(\pi \tau t_1)^{1/2}} \sim \left(\frac{v_S \Delta t}{l}\right) \Delta T_0, \quad (33)$$

which is proportional to the ration of the volume of the sample initially heated by the temperature source to the total volume.

In the diffusion limit $\tau \to 0$, $t \gg t_0$ the asymptotic form of eq.(29) is

$$T(l,t) = \sum_{i=1}^{N} \frac{\Delta T_0 \Delta t}{(4\pi \tau t)^{1/2}} \exp(-t_i^2/4\tau t), \quad (34)$$

where the expansion of $z_i$ in eq.(29) is only valid for $t \gg t_i$. Since the $i$th term contributes to $T(l,t)$ only for $t \sim t_i^2/\tau$, it is valid to use the asymptotic expansion of each term. Eq.(34) is the solution of the diffusion equation (19) with $\tau = 0$, i.e.

$$\frac{\partial T}{\partial t} = \tau v_S^2 \nabla^2 T. \quad (35)$$

This equation may also be solved in the form

$$T(x,t) = \frac{1}{l} \int_0^l T(x,0)dx + \frac{2}{l} \sum_{n=1}^{\infty} \exp(-n^2\pi^2 \tau t/t_0^2) \times \cos\frac{n\pi x}{l} \int_0^l T(x',0)\cos\frac{n\pi x'}{l} dx', \quad (36)$$

where we have used the set of functions which naturally incorporate the boundary condition at the sample ends. For an initial temperature distribution such as (25) one finds [6]:

$$T(l,t) = \Delta T_0 \frac{v_S \Delta t}{l} \left[1 + 2\sum_{n=1}^{\infty} (-1)^n \exp(-n^2\pi^2 \tau t/t_0^2)\right]. \quad (37)$$

Note the prefactor is just the final temperature given by eq.(33). In this form it is particularly convenient to see how to use measurements of $T(l,t)$ to learn the relaxation time $\tau$. At long times, eq.(38) is asymptotically:

$$T(l,t) \sim T_f [1 - 2\exp(-\pi \tau t/t_0^2)] \quad (38)$$

So that $T_f(l) - T(l,t)$ may be used to measure $\tau$.

4. Conclusions

Following method presented in [1] we obtain the equation which describes the propagation of a second sound in nuclear matter. The obtained equation incorporates in natural way the finite velocity of heat propagation. The proposed equation is obtained as a consequence of non-linear heat transfer and is not the result of adding the second derivative term to the Fourier equation In our case that new term describes memory of the nuclear system.

The solution of the new equation describes, in addition to the well-known heat diffusion, also the ballistic propagation of the temperature pulse.